\documentclass[%
 aip,
 amsmath,amssymb,
 reprint,%
]{revtex4-2}

\usepackage{color} 
\usepackage{graphicx}
\usepackage{dcolumn}
\usepackage{bm}

\usepackage{siunitx}

\usepackage[utf8]{inputenc}
\usepackage[T1]{fontenc}
\usepackage{mathptmx}
\usepackage{hyperref}

\draft 

\begin{document}


\title{Cross-plane thermal conductivity of GaN/AlN superlattices} 
\date{\today}



\author{Anna Spindlberger}
  \altaffiliation{Authors to whom correspondence should be addressed: anna.spindlberger@jku.at and alberta.bonanni@jku.at}
\author{Dmytro Kysylychyn} 
\author{Lukas Thumfart} 
\author{Rajdeep Adhikari} 
\affiliation{Institute of Semiconductor and Solid-State Physics, Johannes Kepler University Linz, Altenbergerstr. 69, 4040 Linz, Austria}%
\author{Armando Rastelli} 
\author{Alberta Bonanni}
  \altaffiliation{Authors to whom correspondence should be addressed: anna.spindlberger@jku.at and alberta.bonanni@jku.at}
\affiliation{Institute of Semiconductor and Solid-State Physics, Johannes Kepler University Linz, Altenbergerstr. 69, 4040 Linz, Austria}%


\begin{abstract}
Heterostructures consisting of alternating GaN/AlN epitaxial layers represent the building-blocks of state-of-the-art devices employed for active cooling and energy-saving lightning. Insights into the heat conduction of these structures are essential in the perspective of improving the heat management for prospective applications. Here, the cross-plane (perpendicular to the sample's surface) thermal conductivity of GaN/AlN superlattices as a function of the layers' thickness is established by employing the $3\omega$-method. Moreover, the role of interdiffusion at the interfaces on the phonon scattering is taken into account in the modelling and data treatment. It is found, that the cross-plane thermal conductivity of the epitaxial heterostructures can be driven to values as low as \SI{5.9}{W/(m\cdot K)} comparable with those reported for amorphous films, thus opening wide perspectives for optimized heat management in III-nitride-based epitaxial multilayers.

\end{abstract}


\maketitle 

The tuneable bandgap, high thermal stability, high electric field strength, high electron mobility, and the presence of a spontaneous and piezoelectric polarization in wurtzite III-nitrides led to the integration of GaN and its alloys in state-of-the-art electronics technology\cite{chen_Mater.Sci.Eng.RRep_2020}. Current III-nitride based devices range from light emitting diodes in the ultraviolet\cite{taniyasu_Appl.Phys.Lett._2011,kamiya_Appl.Phys.Lett._2011} and visible range\cite{nakamura_Rev.Mod.Phys._2015, akasaki_Rev.Mod.Phys._2015} to high-electron mobility transistors\cite{dabiran_Appl.Phys.Lett._2008,shen_IEEEElectronDeviceLett._2001,adhikari_Appl.Phys.Lett._2016} and biosensors\cite{li_Nanoscale_2017}. Another research field involving III-nitrides is opened up by doping them with transition metals and by studying the emergent spin related phenomena\cite{dietl_J.Semicond._2019, dietl_Rev.Mod.Phys._2015,devillers_CrystalGrowth&Design_2015,adhikari_Phys.Rev.B_2016}. 
Generally, devices experience efficiency losses due to internal heating\cite{wang_IEEETrans.ElectronDevices_2012,vitusevich_Appl.Phys.Lett._2003} and low thermal conductivity materials are needed as the basis for thermoelectric devices\cite{cahill_JournalofAppliedPhysics_2002, snyder_Nat.Mater._2008} or for on-device cooling\cite{fan_Appl.Phys.Lett._2001}. Thus, a detailed understanding of the thermal properties of thin films and nitride based heterostructures is mandatory in the perspective of efficient thermal management.

Previous works have investigated the thermal conductivity of III-nitrides in bulk\cite{inyushkin_JetpLett._2020} and in thin films\cite{daly_J.Appl.Phys._2002} in detail. Furthermore, several groups studied the decrease of thermal conductivity with increasing Al concentration in Al$_x$Ga$_{1-x}$N bulk\cite{liu_Appl.Phys.Lett._2004} and thin films\cite{vermeersch_Appl.Phys.Lett._2016,daly_J.Appl.Phys._2002,liu_J.Appl.Phys._2005,filatova-zalewska_SolidStateSci._2020,zou_J.Appl.Phys._2006}. A decrease of the bulk thermal conductivity of III-nitrides due to impurity doping\cite{paskov_AIPAdvances_2017,xu_J.Appl.Phys._2019,park_J.Appl.Phys._2019,katre_Phys.Rev.Materials_2018} was also found, while an enhanced conductivity is reported for isotopically enriched GaN\cite{zheng_Phys.Rev.Materials_2019}. For AlN/GaN stacks, significant changes in the phonon properties due to interface effects\cite{zhang_Appl.Phys.A_2019,zhu_AIPAdvances_2019,yang_ChinesePhys.B_2019,polanco_Phys.Rev.B_2019} and strain\cite{tang_J.Appl.Phys._2020} were reported, while for compositionally graded interfaces it was found that alloy scattering is dominant over mismatch scattering\cite{vanroekeghem_Phys.Rev.Applied_2019} of phonons. 


In order to minimize the thermal conductivity of a heterostructure, either the thickness of the single layers must be reduced or scattering centers are introduced by interdiffusion. For instance, in the case of Al$_{0.4}$Ga$_{0.6}$N, the cross-plane (perpendicular to the sample's surface) thermal conductivity $\kappa_{\perp}$ is theoretically predicted to decrease to \SI{1}{W/(m\cdot K)}\cite{vermeersch_Appl.Phys.Lett._2016} when reducing the layer thickness down to \SI{2}{nm}. Due to the anisotropy of the thermal conductivity tensor, the in-plane component $\kappa_{\parallel}$ of the thermal conductivity is expected to exhibit a different heat conduction than the cross-plane one\cite{mei_JournalofAppliedPhysics_2015}. The interface resistance affects the interface crossing of phonons and leads to a quenching of the thermal conductivity, as reported for Ge/Si superlattices\cite{chen_Phys.Rev.Lett._2013,thumfart_J.Phys.D:Appl.Phys._2018,feng_Sci.ChinaPhys.Mech.Astron._2015} and for up to eight pairs of AlN/GaN multilayers\cite{koh_Adv.Funct.Mater._2009}.

\begin{figure}[b]
\includegraphics{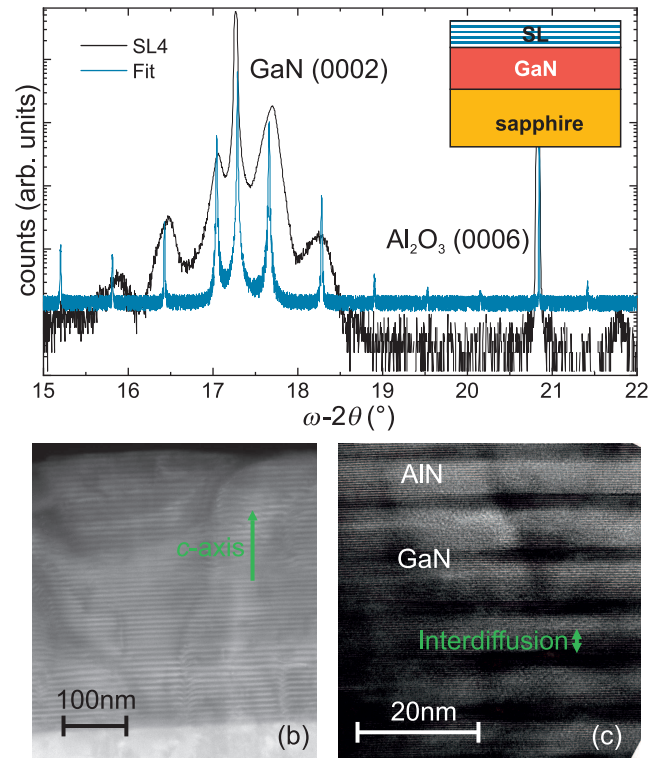}
\caption{\label{fig:structure}(a) XRD scan around the GaN (0002) for SL4 and the fitted signal. Inset: schematic sample structure. HAADF image of the SL4: (b) the whole structure and (c) a magnified image of the interdiffusion region between GaN and AlN.}
\end{figure}

In this work, the thermal conductivity of heterostructures consisting of up to 50  GaN/AlN pairs with different layer thickness is measured by the differential $3\omega-$method\cite{cahill_Rev.Sci.Instrum._1990,lee_J.Appl.Phys._1997} and simulated with \textit{ab-initio} calculations.


The investigated samples are grown by metal organic vapor phase epitaxy (MOVPE) according to the procedure detailed in the supplementary material. In particular, all samples are grown on a $c$-sapphire substrate, upon deposition of a low temperature nucleation layer followed by a \SI{1.5}{\micro m} thick GaN buffer layer and a GaN/AlN superlattice (SL). The considered samples, together with their relevant parameters are listed in Table~\ref{tab:samples}. The SL heterostructures consist of 50 GaN/AlN pairs. The thicknesses of each GaN and AlN layer is varied between \SI{4}{nm} and \SI{16}{nm} over the sample series, covering typical thicknesses for GaN/AlN quantum wells\cite{kaminska_JournalofAppliedPhysics_2016} and the number of layers needed for high-reflectivity bands of distributed Bragg mirrors\cite{zhang_Appl.Sci._2019,shan_ACSPhotonics_2019}. A sketch of the SL structure is given in the inset to FIG.~\ref{fig:structure}~(a) and in the following the SLs are indicated by SL4, SL8, SL12, SL16, where the numbers refer to  the thickness of the single layers in \si{nm}. Additionally, two reference samples, namely a clean sapphire substrate and a GaN buffer, are also included in the study.

The crystallographic properties are analysed by high-resolution x-ray diffraction (XRD) and transmission electron microscopy (TEM). In FIG.~\ref{fig:structure}~(a), the $\omega-2\theta$ scan of the SL4 around the GaN (0002) reflection is plotted and the signal simulated with the AMASS software from PANalytical is given. The satellite peaks of the main GaN reflection arise due to the interference in the diffraction of the different layers of the multi-layer structure. The individual layers of the SL are evidenced in the high-angle annular dark-field image (HAADF), given for sample SL4 in FIG.~\ref{fig:structure}~(b). The upward arrow indicates the growth direction.
Due to interdiffusion, the interfaces between the GaN and the AlN layers are smeared out in an Al$_x$Ga$_{1-x}$N layer\cite{gusenbauer_Phys.Rev.B_2011}, as evidenced by the EDX measurements shown in FIGs.~S3~(a) and (b) of the supplementary material and marked by the arrow in FIG.~\ref{fig:structure}~(c). The interdiffusion at the interface can suppress the coherent nature of phonons, favours diffusive scattering and influences the cross-plane heat conduction\cite{mei_JournalofAppliedPhysics_2015,koh_Adv.Funct.Mater._2009, garg_Phys.Rev.B_2013,huberman_Phys.Rev.B_2013}. The detailed analysis of the layers' thickness with TEM and XRD can be found in the supplementary material, where the TEM images in FIG.~S2~(a)-(d) display V-shaped defects also reported in previous works\cite{li_Jpn.J.Appl.Phys._2002}. 

For the thermal conductivity measurements, a \SI{60}{nm} thin insulating AlO$_x$ layer is deposited onto the specimens, followed by a metal structure consisting of a \SI{120}{nm}-thick and \SI{10}{\micro m} wide Au layer, which is used for heating the underlying material and for determining the corresponding temperature rise. The fabrication steps and geometry of the contacts are given in the section~II.C of the supplementary material.

\begin{figure}[b]
\includegraphics[width=0.45\textwidth]{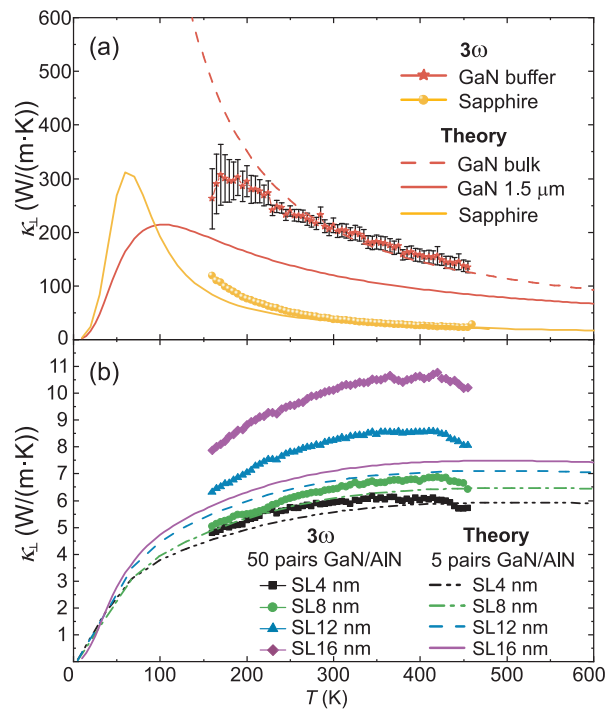}
\caption{\label{fig:thermal_conductivity}(a) Cross-plane thermal conductivity $\kappa_{\perp}$ of the reference samples, sapphire substrate and substrate + GaN buffer, and (b) of the 4 SLs as measured with the differential $3\omega$-method (symbols) and simulated (line).}
\end{figure}


\begin{table}[]
\caption{\label{tab:samples}Samples' details including layer thickness, effective thermal conductivity $\kappa_{3\omega}$ and simulated thermal conductivity $\kappa_\text{sim}$ at \SI{300}{K}.}
\begin{ruledtabular}
\begin{tabular}{lc|cc}
sample & layer thickness (\si{nm}) & $\kappa_{3\omega}$ (\si{W/(m\cdot K)}) & $\kappa_{\text{sim}}$ (\si{W/(m\cdot K)})\\\hline
sapphire & 330000 & 38.0 & 34.6\\
GaN & 1500 & 213.4 & 120.1\\\hline
SL4 & 4/4 & 5.9\footnote{SL with 50 GaN/AlN pairs} & 5.6\footnote{simulated SL with five GaN/AlN pairs}\\
SL8 & 8/8 & 6.4\footnotemark[1] & 6.1\footnotemark[2]\\ 
SL12 & 12/12 & 8.2\footnotemark[1] & 6.7\footnotemark[2]\\
SL16 & 16/16 & 10.1\footnotemark[1] & 7.1\footnotemark[2]\\
\end{tabular}
\end{ruledtabular}
\end{table}

Measurements of the cross-plane thermal conductivity $\kappa_{\perp}$ are performed with the differential $3\omega$-method\cite{cahill_Rev.Sci.Instrum._1990} in the temperature range between \SI{160}{K} and \SI{455}{K}. First the thermal conductivity $\kappa_{\perp}$ of the sapphire substrate and of the GaN buffer layer are measured. The obtained values serve as references for the thermal conductivity for the SLs, as their influence on the measurement have to be deducted. The data evaluation is carried out by solving the heat diffusion equation in a matrix formalism and by fitting the measured temperature oscillations as a function of the heater frequency\cite{tong_ReviewofScientificInstruments_2006}. The whole SL is treated as a single layer.

The measured cross-plane thermal conductivity of the clean substrate and of the buffer deposited onto the substrate as a function of temperature $T$ in the interval \SIrange[range-phrase = -- ,range-units = brackets]{160}{455}{K} are plotted as symbols in FIG.~\ref{fig:thermal_conductivity}~(a) and the values for room temperature are included in Table~\ref{tab:samples}. The simulated conductivity \textit{versus} temperature of the reference samples is also given in FIG.~\ref{fig:thermal_conductivity}~(a). The theoretical calculations for the sapphire substrate are performed following the procedure by Burghartz \textit{et al.}\cite{burghartz_J.Nucl.Mater._1994} with a sapphire thickness of \SI{330}{\micro m} and a temperature dependent specific heat capacity taken from Ref. [\!\citenum{archer_JournalofPhysicalandChemicalReferenceData_1993}]. The values of the measured sapphire substrate follow the trend of the simulated conductivity up to \SI{200}{K}. The measured GaN film thermal conductivity slightly increases for decreasing temperatures and does not feature the steep slope below \SI{200}{K}, that would be expected for bulk GaN. 

The measured cross-plane thermal conductivity as a function of $T$ of the four considered samples containing SLs is reported in FIG.~\ref{fig:thermal_conductivity}~(b) and the obtained values at room temperature are collected in Table~\ref{tab:samples}. For the SL4 with the thinnest individual layers also the lowest thermal conductivity of \SI{5.9}{W/(m\cdot K)} at \SI{300}{K} is measured, while for the SL16 the thermal conductivity reaches \SI{10.1}{W/(m\cdot K)} at the same temperature. The measured signal is mainly influenced by the SL and by the substrate, while the contribution of the buffer layer is negligible for the investigated samples, leading to an error on the measurements lower than \SI{1.5}{\%} over the whole temperature range. A decrease in thermal conductivity with decreasing temperature is found. 
In addition to the experiments, theoretical simulation accounting for phonon scattering mechanisms are performed and included as lines in FIG.~\ref{fig:thermal_conductivity}~(b).

\begin{figure}
\includegraphics{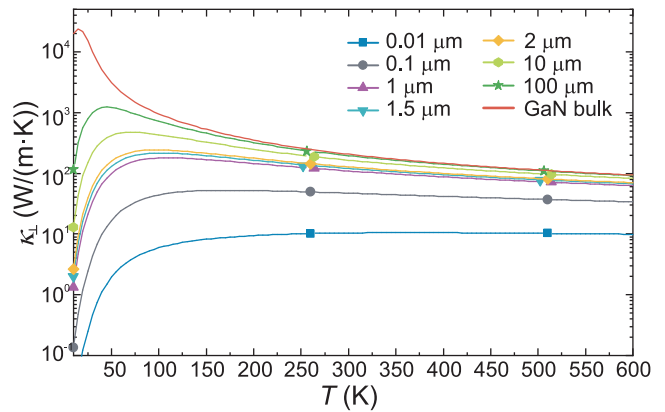}
\caption{\label{fig:sim} Simulated bulk and thin film cross-plane thermal conductivity of GaN as a function of temperature $T$.}
\end{figure}

The calculations of the bulk GaN and GaN/AlN SL thermal conductivity are performed with the \texttt{almaBTE} software package\cite{carrete_Comput.Phys.Commun._2017} in the relaxation time approximation (RTA) and with an uniform wavevector grid of $24^3$. Further details about the program can be found in section~III of the supplementary material. First, the temperature dependent thermal conductivity parallel to the $c$-axis of GaN is analysed. The simulated results are plotted in FIG.~\ref{fig:sim}, where the topmost curve is the calculated conductivity for bulk GaN. The same simulation results are included for comparison in FIG.~\ref{fig:thermal_conductivity}~(a)  (solid line). The simulations are repeated for several layer thicknesses ranging from \SIrange{0.01}{100}{\micro m}. The thickness of the MOVPE grown GaN buffer is \SI{1.5}{\micro m}, therefore a simulation of a GaN layer with the same thickness is included as line combined with reverse triangles in FIG.~\ref{fig:sim} and as dashed line in FIG.~\ref{fig:thermal_conductivity}~(a). The $\kappa_{\perp}$ conductivity of GaN reaches a maximum at \SI{100}{K} for a film thickness of \SI{1.5}{\micro m}, and for thicker layers this maximum is shifted to lower temperatures, namely around \SI{15}{K} for bulk GaN. Inyushkin \textit{et al.}\cite{inyushkin_JetpLett._2020} reported a maximum of \SI{3770}{W/(m\cdot K)} around \SI{28}{K}. No sharp maximum is reached for films with thicknesses of \SI{1}{\micro m} and lower, for which, instead, the conductivity saturates at temperatures above \SI{100}{K}. The highest reached thermal conductivity decreases with decreasing layer thickness, due to the significant effect of interface- and boundary-scattering of phonons.

Two limit regimes of heat conduction are relevant for crystalline films. In particular, in films with thicknesses much lower than the cross-plane mean free path $\lambda_{\text{dom}}$\cite{vermeersch_Appl.Phys.Lett._2016} of phonons, which depends on the cross-plane bulk thermal conductivity and on the mean free path of the cross-plane projection, $\kappa_{\perp}$ is predominantly ballistic. In contrast, for films with film thickness much larger than $\lambda_{\text{dom}}$, the heat conduction is quasi-diffusive and therefore no distinct dependence on the film thickness is found. The film thicknesses considered in this work are smaller than the characteristic lengths\cite{vermeersch_Appl.Phys.Lett._2016} for the transition to the ballistic regime, thus a distinct change of $\kappa_{\perp}$ with different individual layer thicknesses is expected.

\begin{figure}[b]
\includegraphics{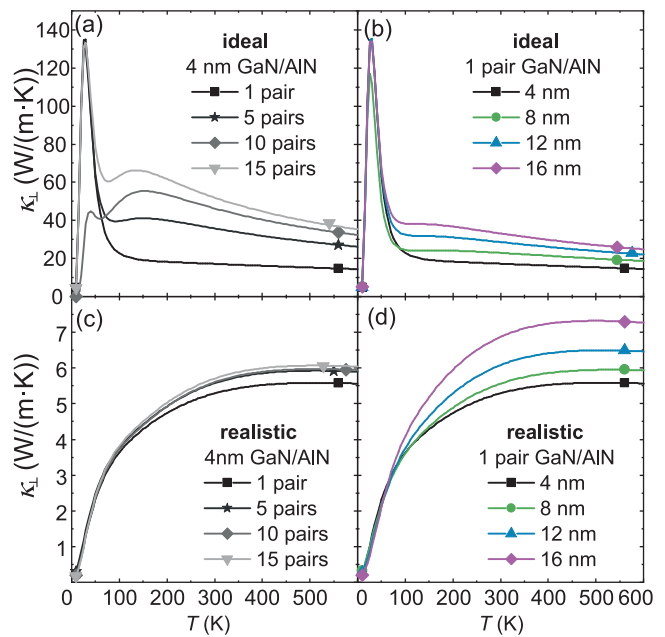}
\caption{\label{fig:sim-SL} Simulated temperature dependent thermal conductivity of an ideal GaN/AlN superlattice for (a) different number of pairs with \SI{4}{nm} and (b) several layer thicknesses with one pair. Temperature dependent simulation of realistic SLs (c) with individual layer thickness of \SI{4}{nm} with different number of pairs over the series and (d) one single pair with different thicknesses over the samples' series.}
\end{figure}

In order to gain further insights into the effect of alloy scattering, first a heterostructure is simulated, which assumes a well defined stoichiometry of the GaN/AlN layers without intermixing and without a substrate. This ideal structure has a concentration $x$ of Al of either $0$ or $1$. The SL crystal structure is built with the superlattice package of the \texttt{almaBTE} software, following the procedure reported in Ref.~[\!\citenum{chen_Phys.Rev.Lett._2013}]. In FIG.~\ref{fig:sim-SL}~(a), the computed temperature dependent cross-plane thermal conductivity of the SL4 for one to 15 pairs of GaN/AlN is given. The thermal conductivity increases over the whole temperature range with increasing number of SL pairs. For completeness, also an inverse structure AlN/GaN (not shown), is calculated for five pairs, and the same thermal conductivity as in the case of the GaN/AlN sequence is found. In FIG.~\ref{fig:sim-SL}~(b), the simulated $\kappa_{\perp}$ of one SL pair for all four thicknesses of the individual layer, \textit{i.e.} \SI{4}{nm}, \SI{8}{nm}, \SI{12}{nm} and \SI{16}{nm}, are reported over temperature. The effect of scattering of phonons at interfaces is less pronounced with increasing layer thickness, leading to an increase in thermal conductivity\cite{mizuno_Sci.Rep._2015,chen_Phys.Rev.B_1998}, as evidenced in FIG.~\ref{fig:sim-SL}~(b). 

In order to account for interdiffusion at the interfaces, a concentration-graded layer of Al$_x$Ga$_{1-x}$N between the two ideal AlN and GaN layers is introduced in the calculations (realistic structures), while the total thickness of the structure is kept constant. In FIG.~\ref{fig:sim-SL}~(c), $\kappa_{\perp}$ over the temperature is given for SL4 with different number of pairs over the samples' series. When comparing the simulated ideal structures with those including Al$_x$Ga$_{1-x}$N regions, a significant decrease in thermal conductivity is found. In the realistic structures, additionally to the phonon interface scattering, alloy scattering is observed. As reported by Vermeersch \textit{et al.}\cite{vermeersch_Appl.Phys.Lett._2016} the change of Al concentration from GaN towards AlN leads to a decrease in the thermal conductivity of up to one order of magnitude in bulk materials. The increase in the number of layers slightly increases the thermal conductivity of the whole structure, in accord to Ref.~[\!\citenum{carrete_J.Phys.Chem.C_2018}]. In FIG.~\ref{fig:sim-SL}~(d) the conductivity of one GaN/AlN pair with intermixing for the four thicknesses is plotted over temperature. As found in the case of ideal SLs, the increase of an individual layer thickness induces an increment of the thermal conductivity. The thermal conductivity of the realistic SLs is less influenced by a change in layer numbers than by an increase of individual layer thickness.\\
The simulations of five realistic GaN/AlN pairs qualitatively reproduce the shape of the measured $\kappa_{\perp}$ given in FIG.~\ref{fig:thermal_conductivity}~(b). As discussed, a slight increase in the maximum value of the thermal conductivity is expected when increasing the number of pairs from five to 50. Additional changes in the thermal conductivity can arise by a variance in the Al concentration and increases when going towards an ideal SL. The simulations do not consider any kind of defects, though the V-shaped ones, present in the MOVPE grown samples, are expected to reduce the thermal conductivity. Additionally, the simulations do not contain a substrate, which is expected to influence the phonon transport at the substrate and SL boundary. Furthermore, the influence of strain, predicted to increase the thermal conductivity\cite{zhu_AIPAdvances_2019}, in the layers is not treated in the simulations. 

\begin{figure}
\includegraphics{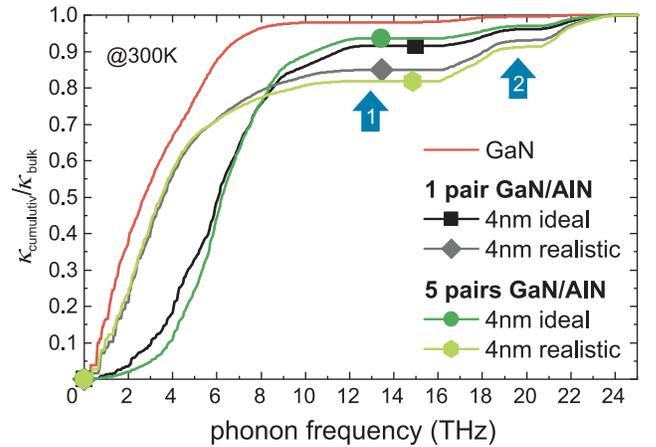}
\caption{\label{fig:PhononF}Normalized cumulative thermal conductivity over the contributing phonon frequencies for realistic and ideal SLs.}
\end{figure}

In semiconductors, thermal energy is mainly transported by phonons and the phonon frequency spectra can be used to identify the main scattering mechanism. In FIG.~\ref{fig:PhononF}, the fraction of the normalized cumulative thermal conductivity, obtained by dividing the cumulative thermal conductivity by the bulk thermal conductivity transferred by phonons as a function of the frequency is given at \SI{300}{K}. The simulated phonon contribution for a bulk GaN is plotted as a solid line. Moreover, the four lines with symbols represent the simulations for one pair of GaN/AlN with ideal and realistic composition and five pairs of GaN/AlN with ideal and realistic composition, respectively. For realistic SLs, a major contribution to the heat conductivity originates from phonons with frequency between \SI{0}{THz} and \SI{6}{THz}, while phonons between \SI{12}{THz} and \SI{16}{THz} contribute minimally, leading to a first plateau marked by the upward arrow labeled with 1 in FIG.~\ref{fig:PhononF}. A second plateau occurs above \SI{19}{THz}, indicated by the upward arrow two, and phonons with frequencies higher than \SI{22}{THz} do not contribute significantly to the thermal conductivity. The splitting in two plateaus is due to the alloy scattering of high-frequency phonons\cite{chen_Phys.Rev.Lett._2013}. Due to an enhancement of low-frequency phonon scattering at the abrupt interfaces\cite{chen_Phys.Rev.Lett._2013}, the conductivity of the ideal SLs is less significantly affected by phonons in the frequency range \SIrange[range-phrase = --,range-units = brackets]{0}{4}{THz}. Only one prominent plateau above \SI{8}{THz} is visible, due to the missing effect of alloy scattering. For completeness, the simulated cumulative thermal conductivity normalized to the bulk is provided as a function of the mean free path (MFP) of the phonons for bulk AlN, bulk GaN, and - in both ideal and realistic case - for two superlattices with one and five GaN/AlN pairs, respectively, in FIG.~S4 of the supplementary material.

In conclusion, the cross-plane thermal conductivity of sapphire, GaN, and SLs consisting of up to 50 pairs of GaN/AlN has been investigated. The temperature dependent conductivity of various film thicknesses of GaN has been simulated and the values are reproduced by measuring the thermal conductivity with the differential $3\omega$-method. The simulated thermal conductivity of the GaN/AlN SL with different number of pairs has been compared and it is found that the change in absolute values is minimal. The measured values for SLs with individual layer thicknesses of \SIrange{4}{16}{nm} and 50 pairs increases at room temperature from \SIrange{5.9}{10.1}{W/(m\cdot K)}. These results confirm the possibility of decreasing the cross-plane thermal conductivity of epitaxial crystalline GaN/AlN heterostructures to values as low as those reported for amorphous layers\cite{mizuno_Sci.Rep._2015}, opening wide perspectives for heat management in III-nitride-based devices.\\

See the supplementary material for the details about the epitaxial growth, the in depth TEM and XRD characterization and the simulations.

This work was supported by the European Commission’s Horizon 2020 Research and Innovation Program [Grant No. 645776 (ALMA)] and by the Austrian Science Fund (FWF) [Project P31423]. The authors thank Werner Ginzinger for preparing the specimens for the TEM experiments and for carrying out the related measurements, Heiko Groiss and Jesús Carrete for fruitful discussions, and Albin Schwarz for preparing the contact structures for the $3\omega$ measurements.

\section*{data availability}
The data that support the findings of this study are available from the corresponding author upon reasonable request.\\

This article may be downloaded for personal use only. Any other use requires prior permission of the author and AIP Publishing. This article appeared in A. Spindlberger \textit{et al.}, Appl. Phys. Lett. 118, 062105 (2021) and may be found at \url{https://doi.org/10.1063/5.0040811}

\bibliography{Thermal}

\end{document}



\title{Cross-plane thermal conductivity of GaN/AlN superlattices:\\
Supplementary Material} 
\date{\today}



\author{Anna Spindlberger}
  \altaffiliation{Authors to whom correspondence should be addressed: anna.spindlberger@jku.at and alberta.bonanni@jku.at}
\author{Dmytro Kysylychyn}
\author{Lukas Thumfart}
\author{Rajdeep Adhikari}
\affiliation{Institute of Semiconductor and Solid-State Physics, Johannes Kepler University Linz, Altenbergerstr. 69, 4040 Linz, Austria}%
\author{Armando Rastelli}
\author{Alberta Bonanni}
  \altaffiliation{Authors to whom correspondence should be addressed: anna.spindlberger@jku.at and alberta.bonanni@jku.at}
\affiliation{Institute of Semiconductor and Solid-State Physics, Johannes Kepler University Linz, Altenbergerstr. 69, 4040 Linz, Austria}%



\maketitle 

\beginsupplement
\section{Epitaxial growth}
The investigated samples are fabricated in an AIXTRON 200RF horizontal tube metal organic vapor phase epitaxy (MOVPE) reactor on $c$-plane sapphire substrates. As precursors TMGa, TMAl and NH$_3$ for Ga, Al and N respectively are used and H$_2$ as carrier gas. The flow rates are kept constant at \si{25} standard cubic centimeters per minute (sccm) for Ga and Al and \SI{1500}{sccm} for NH$_3$ at the whole growth process. First the substrate is heated to \SI{1080}{\degree C} and floated with ammonia for nitridation, then a thin GaN nucleation layer is grown at \SI{540}{\degree C} and annealed at \SI{1040}{\degree C}. The growth continues with a \SI{1.5}{\micro m} thick buffer layer of GaN and for the GaN/AlN hetero-structure 50 pairs of GaN on AlN layers are deposited. For this study, the thicknesses of the individual superlattice (SL) layers varies from \SIrange{4}{16}{nm} over the samples' series. The growth process is monitored \textit{via} \textit{in-situ} reflectometry.

\section{Details on Characterization}
\subsection{X-ray diffraction}

\begin{figure*}
\includegraphics{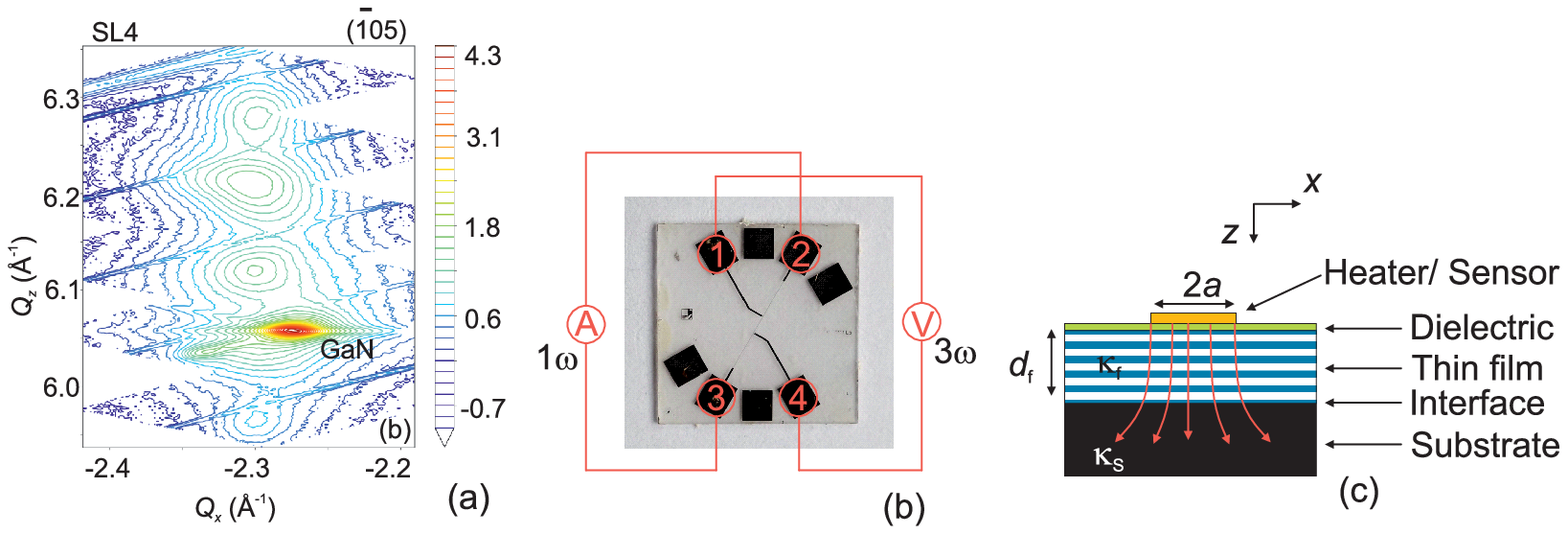}
\caption{\label{fig:XRD_3omega}(a) RSM around the $(\overline{1}05)$ peak of GaN of SL4. (b) Geometry used for the differential $3\omega$ measurements with contact pads for current $I$ and voltage $V$ (c) Schematic sample structure of the SL with overall thermal film conductivity $\kappa_f$ and substrate conductivity $\kappa_s$}
\end{figure*}

High-resolution x-ray diffraction (HRXRD) has been carried out with a PANalytical's X'Pert PRO Materials Research Diffractometer (MRD) equipped with a hybrid monochromator with a \SI{1/4}{\degree} divergence slit. The MRD includes a PixCel detector with 256-channels and a \SI{11.2}{mm} anti-scatter slit.\\
To gain insights in the relaxation of the samples, reciprocal space maps (RSM) are performed at the $(\overline{1}05)$ diffraction peak of GaN. The most prominant peak in the RSM in FIG.~\ref{fig:XRD_3omega}~(a) for SL4 is due to the GaN buffer. The fringes around this peak, are due to the 50 pairs of GaN/AlN and the SL in all considered samples exhibit $\sim\SI{50}{\%}$ relaxation on to the GaN buffer layer (different $Q_x$ value for the peaks of the fringes).\\
Moreover, $\omega-2\theta$ scans are performed, where the thickness fringes of the SL can be seen. The fringes are reproduced by simulating the $\omega-2\theta$ signal of the SL with the AMASS software from PANalytical and fitting the measured signal. Due to interdiffusion between the AlN and GaN layer, a graded Al$_x$Ga$_{1-x}$N layer of thickness $d_\text{AlGaN}$ between the two SL layers is considered in the simulation. The fitted thicknesses and Al concentrations of each layer are summarized in Table \ref{tab:XRD_simulation} and an example simulation is given in FIG.~1~(a) of the main text. The measurement exhibit broader peaks than the simulation.

\begin{table}
\caption{\label{tab:XRD_simulation}Fitted layer thicknesses of the SL layers and the interdiffusion region ($d_\text{AlGaN}$).}
\begin{ruledtabular}
\begin{tabular}{lccc}
sample & $d_{\text{GaN}}$ (\si{nm}) & $d_{\text{AlN}}$ (\si{nm}) & $d_{\text{AlGaN}}$ (\si{nm}) \\\hline
SL4 & 4.09 &  2.22 & 1.18 \\
SL8 & 10.00 &  0.82 & 5.01 \\
SL12 & 13.88 & 3.99 & 8.68 \\
\end{tabular}
\end{ruledtabular}
\end{table}

\subsection{Transmission Electron Microscopy}
The transmission electron microscope (TEM) used in this study is a JEOL JEM-2200FS TEM microscope operated at \SI{200}{kV} in conventional and high-resolution imaging (HRTEM) modes. The preparation of the TEM specimens, both cross-section and plan-view, is done by a conventional procedure of mechanical polishing followed by Ar+ milling and double-side plasma cleaning. For the elemental analysis, energy dispersive x-ray spectroscopy (EDX) is performed while measuring the samples in scanning TEM mode (STEM). To gain a broad overview of the sample structure high-angle annular dark-field imaging (HAADF) is also conducted.\\

\begin{figure}
\includegraphics{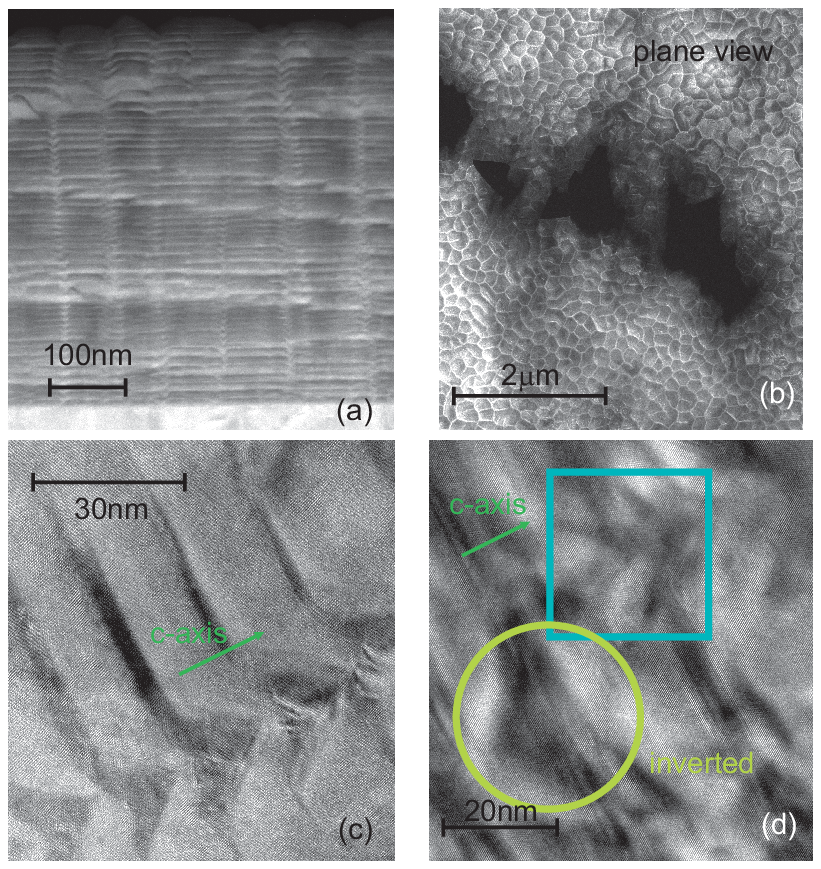}
\caption{\label{fig:TEM}(a) HAADF image of SL12. (b) Plane view of SL16 (c) Magnification of a region with V shape defects of SL12 and (d) with inverted V shape.}
\end{figure}

\begin{figure}
\includegraphics{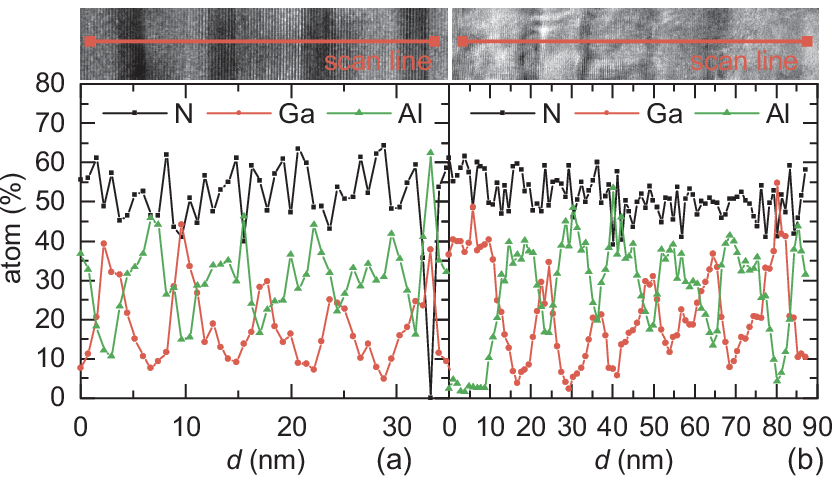}
\caption{\label{fig:EDX}Top panels: cross-section TEM of (a) SL4 and (b) SL8; lower panels: EDX of (a) SL4 and (b) SL8 highlighting the distribution of Ga, N and Al across the heterostructures and in particular the presence of interdiffusion (Al$_x$Ga$_{1-x}$N) between the GaN and AlN layers.}
\end{figure}

In the HAADF images, the individual layers of the SL are visible, as in FIG.~\ref{fig:TEM}~(a) for SL12. The sample contains defects extending from the GaN buffer to the surface. By comparing SL12 to the SL4 given in FIG.~1~(b) of the main text, an increase in defect density is found for the SL12. These defects are most prominent in the SL16 (not shown). FIG.~\ref{fig:TEM}~(c) gives a detailed view of one those V-shaped defects\cite{li_Jpn.J.Appl.Phys._2002}. In the V-pits GaN is preferably accumulated. In the investigated structures inverted V-pits (anti-parallel to the growth direction) are found, as for example in FIG.~\ref{fig:TEM}~(d). In FIG.~\ref{fig:TEM}~(b), a HAADF plane view of SL16 is provided.\\
The EDX measurements shown in FIG.~\ref{fig:EDX}~(a) for SL4 and (b) for SL8 confirm the alternate presence of Ga and Al in the SL layers. Between the AlN and GaN an interdiffusion region is detected.\\


\subsection{Thermal conductivity with the $3\omega$-method\label{sec:3omega}}
The cross-plane thermal conductivity $\kappa_{\perp}$ is measured with the differential $3\omega-$method\cite{cahill_Rev.Sci.Instrum._1990,lee_J.Appl.Phys._1997} for temperatures between \SIrange{160}{455}{K}. For this method, a \SI{60}{nm} thick AlO$_x$ is deposited by atomic layer deposition, followed by several lithography steps to define the heater geometry shown in FIG.~\ref{fig:XRD_3omega}~(b). The heater geometry consists of a \SI{5}{nm} thin Cr adhesion layer followed by a \SI{120}{nm} thick Au layer, fabricated by metal deposition. The strip width is chosen to be $2a= \SI{10}{\micro m}$ and this strip acts as both heater and thermometer. The heater thickness is confirmed by atomic force microscopy and included in the evaluation.\\
The measurements are performed in a frequency range between \SI{40}{Hz} and \SI{4000}{Hz} and a power per unit length of $\SI{15}{W/m}$.


The values of the thermal conductivity are obtained by fitting the measured data and by solving the diffusion equation,\textit{i.e.} by calculating the two dimensional heat flow of a multilayer structure with the transfer matrix method. The fitting routine first treats the sapphire substrate and the buffer layer and than the SL is included as an additional thin film. 

\section{Theoretical approach \label{sec:theory}}
\subsection{AlmaBTE software} 
With the \texttt{almaBTE} software package the phonon transport for bulk crystals or alloys, thin films and also multilayered structures can be calculated. The software solves the space- and time-dependent Boltzmann transport equation (BTE) and calculates the thermal conductance, the effective thermal conductivities and heat-current distributions \cite{carrete_Comput.Phys.Commun._2017}.\\
In contrast to Monte Carlo methods, the \texttt{almaBTE} solver starts with a pre-calculated equilibrium component of the material and after building the desired structure calculates the phonon transport and the related quantities. Several materials are already included in the materials catalog and can be downloaded from the homepage \url{www.almabte.eu}. The material files include information about the geometrical description and the second- and third-order derivatives of its potential energy at equilibrium \cite{carrete_Comput.Phys.Commun._2017}.\\
The package includes the temperature independent 2-phonon scattering rates, elastic scattering rates and at each temperature the resulting 3-phonon scattering rates. With the relaxation time approximation each component of the thermal conductivity tensor is obtained in dependence on the mode contribution of the heat capacity, the volumetric heat capacity, the mean free path and the group velocity.\\

Concerning the simulations, for a SL with \SI{4}{nm} individual layer thickness and one pair, the ideal SL contains eight monolayers of GaN and eight monolayers of AlN. Whereas for a realistic SL six monolayers of GaN and AlN are included and intercalated four monolayers of Al$_x$Ga$_{1-x}$N with $x$ ranging from zero to one are included. For higher number of pairs, the procedure is repeated. In the case of realistic SLs the interdiffusion between each GaN and AlN layer is included.

\begin{figure}
\includegraphics{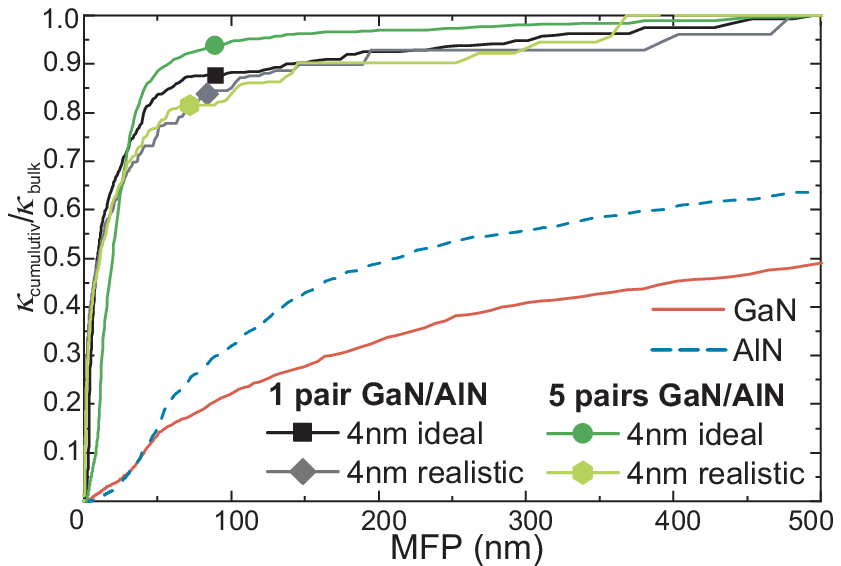}
\caption{\label{fig:MFP}Simulated cumulative thermal conductivity normalized to the bulk over the phonon mean free path for bulk GaN, bulk AlN, and for one pair and five pairs of GaN/AlN in the ideal and realistic case.}
\end{figure}

In FIG.~\ref{fig:MFP} the simulated cumulative thermal conductivity normalized to the bulk is provided as a function of the mean free path (MFP) of the phonons for bulk AlN, bulk GaN, and – in both ideal and realistic case - for two superlattices with one and five GaN/AlN pairs, respectively. For the heterostructures (ideal and realistic), the main contribution to the thermal conductivity is given by phonons with MFP lower than \SI{400}{nm}. In contrast, in the bulk GaN and AlN only half of the total thermal conductivity is transported by phonons with MFP lower than \SI{500}{nm}. It is worth mentioning, that Carrete \textit{et al.}\cite{carrete_Comput.Phys.Commun._2017} reported a MFP up to \SI{10}{\micro m} for the two bulk materials.

\bibliography{Thermal.bib}